\documentclass[reprint,amsmath,amssymb,braket,aps,prb,longbibliography]{revtex4-1} 
\usepackage{amsmath,amssymb}
\usepackage{graphicx,color,braket,amsmath,dcolumn,bm}
\usepackage[colorlinks=true,linkcolor=blue]{hyperref}

\begin{document}

\title{Electron-hole collision-limited resistance of gapped graphene}

\author{Arseny Gribachov}
\affiliation{Moscow Institute of Physics and Technology, Dolgoprudny 141700, Russia}

\author{Vladimir Vyurkov}
\affiliation{Moscow Institute of Physics and Technology, Dolgoprudny 141700, Russia}
\affiliation{Valiev Institute of Physics and Technology RAS, 36/1 Nahimovsky Ave., 117218, Moscow, Russia}

\author{Dmitry Svintsov}
\email{svintcov.da@mipt.ru}
\affiliation{Moscow Institute of Physics and Technology, Dolgoprudny 141700, Russia}

\begin{abstract}
    Collisions between electrons and holes can dominate the carrier scattering in clean graphene samples in the vicinity of charge neutrality point. 
    While electron-hole limited resistance in pristine gapless graphene is well-studied, its evolution with induction of band gap $E_g$ is less explored. Here, we derive the functional dependence of electron-hole limited resistance of gapped graphene $\rho_{eh}$ on the ratio of gap and thermal energy $E_g/kT$. At low temperatures and large band gaps, the resistance grows linearly with $E_g/kT$, and possesses a minimum at $E_g \approx 2.5 kT$. This contrast to the Arrhenius activation-type behaviour for intrinsic semiconductors. Introduction of impurities restores the Arrhenius law for resistivity at low temperatures and/or high doping densities. The hallmark of electron-hole collision effects in graphene resistivity at charge neutrality is the crossover between exponential and power-law resistivity scalings with temperature. 
\end{abstract}
\maketitle

\section{Introduction}
The problem of minimal graphene conductivity $\sigma_{\min}$ observed at charge neutrality has been a subject of long debate since the discovery of graphene. Experimental studies have shown that $\sigma_{\min}$ varies slightly between samples and with changing the temperature~\cite{Novoselov-massless_2deg,DasSarma_MinimumConductivity,Bolotin_TDependent}, which posed questions about the universality of this quantity. Numerous theoretical works attempted to derive $\sigma_{\min}$ from Kubo's formula for clean graphene at zero frequency of electric field $\omega$, zero temperature $T$ and doping $\varepsilon_F$~\cite{Herbut_Kubo1,Mishchenko_Kubo2,Gusynin_Kubo3}. The apparent 'universal' result appeared to depend on the order of taking the limits of zero frequency, temperature and doping~\cite{Ziegler_nonuniversal_Kubo}. The latter fact indicated on the deficiency of model of 'clean graphene' for derivation of universal minimal conductivity.

The first resolution of the minimal conductivity puzzle appeared by realizing that graphene at charge neutrality has some residual doping. This doping comes from random impurities in the sample, arranged in positively and negatively charged clusters. The free carriers of graphene tend to screen these impurity charges, forming the electron-hole puddles~\cite{Klitzing_Puddles}. The root-mean-square density of charge carriers in graphene appears to be non-zero. Instead, it is proportional to the impurity density $n_i$, and so is the carrier scattering rate. These two proportionalities lead to a very weak dependence of $\sigma_{\min}$ on the density of residual impurities and temperature, which can be considered as 'approximate universality'~\cite{Adam-self-consistent-graphene-transport,Falko_RandomResistor}.

With the current level of graphene technology, the density of residual impurities can readily be lower than the density of thermally activated electrons and holes $n_{\rm th} \approx 8\times 10^{10}$ cm$^{-2}$~\cite{Cao_Quality_Heterostructures}. In such a situation, electron-hole puddles and impurity scattering make a minor contribution to the resistivity. The scattering between electrons and holes now governs the experimentally measured value of minimum conductivity~\cite{Morpurgo_NPhys_EHCollisions,Berdyugin_Sci_Schwinger,Bandurin_PRL_EHFriction,Gallagher2019}. A strong violation of Wiedemann-Frantz relation between electrical and thermal conductivity at charge neutrality~\cite{Crossno2016} and the appearance of new electron-hole sound waves~\cite{Zettl_energy_waves} also indicate on the dominant role of e-h scattering in clean samples. A scaling estimate of e-h limited conductivity was presented in Ref.~\onlinecite{Vyurkov_JETP} and resulted in $\sigma_{\min} = C \alpha_c^{-2} e^2/h$, where $\alpha_c = e^2/\hbar v_0$ is the Coulomb coupling constant, $v_0$ is the velocity of massless electrons in graphene, and $C$ is the numerical prefactor. A rigorous solution of the kinetic equation using the variational principle confirmed the result and established $C\approx 0.76$~\cite{Kashuba2008,Fritz2008}.

While most theoretical and experimental studies were devoted to the electron-hole scattering in pristine gapless graphene, very little attention have been paid to the same process in gapped systems. The gap induction in single graphene layer is possible under lattice reconstruction on boron nitride substrates~\cite{Adam_NComms_Gap_hBN}. The gap is readily induced in graphene bilayer under the action of transverse electric fields~\cite{McCann2007}. The derivatives of graphene are not the only examples of 2d electron systems with small energy gap. Another family is represented by quantum wells based on mercury cadmium telluride of sub-critical thickness~\cite{Konig_Science_QSHI,Gavrilenko_HgTe_Dirac}. Given the large variety of clean 2d systems with small band gap and their potential applications in nano- and optoelectronics, it is natural to study the factors limiting their electrical resistivity, particularly, the inevitable electron-hole scattering. An attempt to derive and measure the electron-hole limited resistivity was presented in \onlinecite{Tan_SciAdv_TunableSemimetal}. Its results cannot be considered as satisfactory because the electron-hole scattering times $\tau_{eh}$ in the presence of band gap $E_g$ were not derived, but rather guessed. The authors of  have proposed a universal function $f(E_g/kT)$ governing the scaling of e-h limited conductivity with band gap; the function possessed a quadratic maximum at $E_g = 0$ and dropped exponentially at $E_g/kT \gg 1$. Such dependence was seemingly confirmed by the experiment.

The present paper is aimed at a rigorous derivation of electron-hole scattering time and conductivity at neutrality point in gapped graphene. Our formalism is based on kinetic equation with carrier-carrier collision integral; the carriers are assumed interacting via unscreened Coulomb potential. The kinetic equation is solved with a variational principle which yields good results for the estimates of conductivity~\cite{Fritz2008}. We find that at large band gaps, the conductivity scales as $\sigma_{\min} \propto kT/E_g$. This behaviour differs essentially from conventional Arrhenius-type activation. Such non-Arrhenius behaviour of minimal conductivity can be explained with simple gas kinetics arguments. The free path time of a trial electron against a dilute hole background is inversely proportional to the hole density $\tau_{eh} \propto n^{-1}_h \propto e^{E_g/2kT}$, i.e. grows exponentially with gap induction. The Drude conductivity is proportional to the product of electron density and free-path time, $\sigma_{\min} \propto n_e \tau_{eh}/m^*$. As electron and hole components are balanced at charge neutrality, $n_e = n_h$, the leading Arrhenius exponents are cancelled in the expression for conductivity. Eventually, the conductivity depends on the gap only via effective mass, $m^* = E_g/2v_0^2$. This justifies the hyperbolic dependence of $\sigma_{\min}$ on $E_g$.

The above intuitive explanation is missing the long-range character of Coulomb interaction. In classical plasmas, the latter led to log-divergent collision integral~\cite{Landau_Collisions}. We show that no such divergences appear during the evaluation of conductivity. The collision integral converges both at small momentum transfers $q \rightarrow 0$ as such momenta do not change the electric current, and at large momenta $q \rightarrow \infty$ due to small quantum-mechanical overlap between scattered states. All in all, our variational derivation results in following expression for the conductivity at charge neutrality valid at $E_g \gg kT$
\begin{equation}
    \sigma_{\min} = \frac{8}{\pi}\frac{e^2}{h} \alpha_C^{-2} \frac{kT}{E_g}.
\end{equation}

\section{Variational approach to kinetic equation with carrier-carrier collisions}

Electron states in the gapped graphene are described by a 'massive' Dirac Hamiltonian
$$
\hat{\mathcal{H}}_D = 
\left[
\begin{array}{cc}
 E_g/2 & v_0 \left(\hat{p}_x-i
   \hat{p}_y\right) \\
 v_0 \left({\hat p}_x+i {\hat p}_y\right) &
   -E_g/2 \\
\end{array}
\right]
$$
Such Hamiltonian is applicable to the single layer graphene aligned to boron nitride substrate, and to 2d electron system in CdHgTe quantum wells. Its applicability to graphene bilayer is limited, as the latter has a quadratic band touching at $E_g = 0$. We will further argue that such Hamiltonian is applicable for bilayer at large induced gaps under proper replacement of parameters.  

\begin{figure}[ht]
\center{\includegraphics[width=1\linewidth]{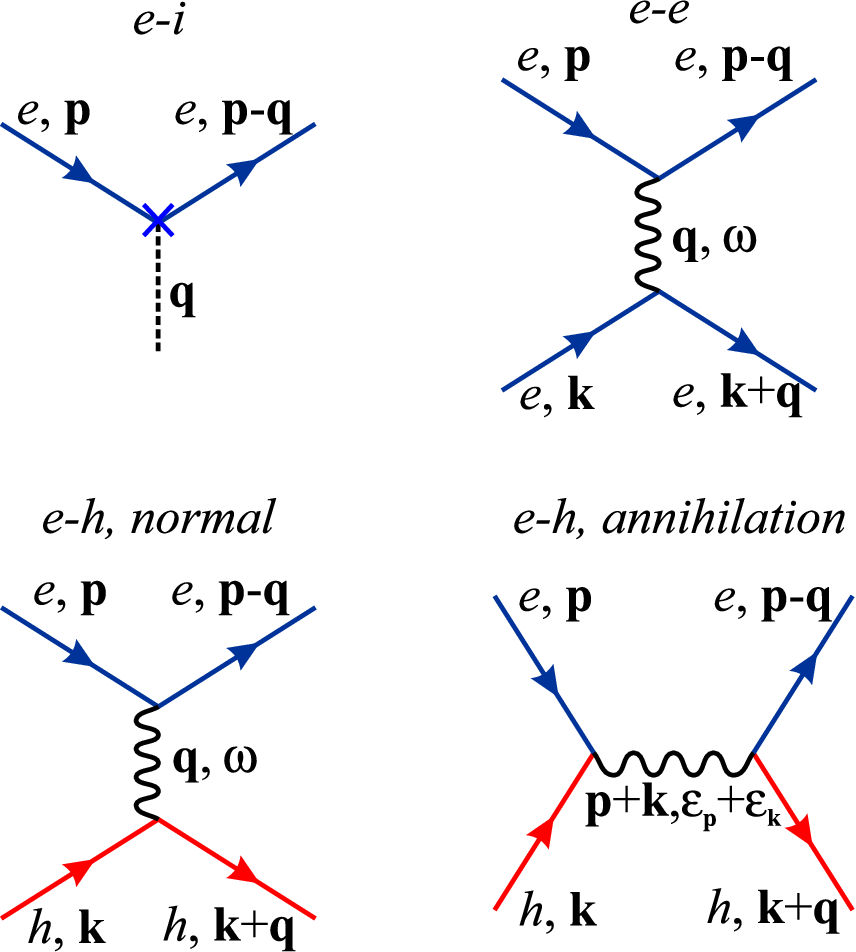}}
\caption{Scattering diagrams for the processes contributing to graphene resistivity at charge neutrality point: electron-impurity, electron-electron, and electron-hole scattering (normal and annihilation-type processes)}
\label{fig-diagrams}
\end{figure}

The dc conductivity is obtained by solving the kinetic equation with respect to distribution function which we linearize as $f_p = f_0 - \Phi_{e/h,\bf p} \partial f_0/\partial \varepsilon$, here $\Phi_{e/h,\bf p}$ is the linear function of electric field ${\bf E}$, the subscripts $e$ and $h$ distinguish between electrons and holes. We restrict our consideration to the carrier-carrier and carrier-impurity collisions. The former involve electron-electron (e-e), hole-hole (h-h), and electron-hole (e-h) collisions (Fig.~\ref{fig-diagrams}). The kinetic equation for electrons takes on the form
\begin{equation}
\label{eq-kinetic}
    - e {\bf E}{\bf v_p} \frac{\partial f_0}{\partial \varepsilon} = -{\mathcal{C}}_{ei}\{\Phi_{e,\bf p}\} - {\mathcal{C}}_{ee}\{\Phi_{e,\bf p}\} - {\mathcal{C}}_{eh}\{\Phi_{e,\bf p},\Phi_{h,\bf p}\},
\end{equation}
and similarly for holes with apparent change of signs. At charge neutrality electrons and holes move in the opposite directions, thus $\Phi_{e,\bf p} = - \Phi_{h,\bf p} \equiv \Phi_{\bf p}$. The complex structure of carrier-carrier collision integrals makes the exact solution of (\ref{eq-kinetic}) impossible, at least in the non-degenerate case~\cite{Combescot_eletron-hole}. For this reason, we use the variational approach with a reasonable trial form of distribution function to get an estimate of the resistivity~\cite{ziman2001electrons,Maldague_PRB_Variational}. The problem of resistivity is weakly sensitive to the specific form of $\Phi_{\bf p}$ (unlike the problems of thermoelectric coefficients~\cite{Takashi_PRB_Variational}), thus the variational approach is efficient for predicting its functional dependence.

Within this approach, one maximizes the entropy generation rate being a quadratic functional of distribution function $\Phi$:
\begin{widetext}

\begin{gather}
\mathcal{Q}[\Phi] = \mathcal{Q}_0[\Phi] + \mathcal{Q}_{ei}[\Phi] + \mathcal{Q}_{ee}[\Phi]+\mathcal{Q}_{eh}[\Phi],\\ 
\mathcal{Q}_0[\Phi] = -e N \sum_{\bf p} 
({\bf v_p} {\bf E} ) \frac{\partial f_0}{\partial \varepsilon} \Phi_{\bf p},\\
{\mathcal{Q}}_{ei}[\Phi] = \frac{N}{2}\sum\limits_{\bf pp'}{\frac{\partial f_0}{\partial \varepsilon} (\Phi_p - \Phi_{p'})^2 W^{ei}_{\bf pp'}},
\end{gather}

\begin{gather}
\label{eq-q-ee}
\mathcal{Q}_{ee}[\Phi] = \frac{N}{8}\frac{1}{kT}\sum\limits_{\bf{pkq}}{{{f}_{0}}\left( \bf{p} \right){{f}_{0}}\left( \bf{k} \right)[1-{{f}_{0}}\left( {\bf{{p}'}} \right)][1-{{f}_{0}}\left( {\bf{{k}'}} \right)]
  W_{\bf{pk}, \bf{{p}'{k}'}}^{ee}{{\left[ \left( {{\Phi }_{p}}+{{\Phi }_{k}} \right)-\left( {{\Phi }_{{{p}'}}}+{{\Phi }_{{{k}'}}} \right) \right]}^{2}} },\\
\label{eq-q-eh}
\mathcal{Q}_{eh}[\Phi] = \frac{N}{8}\frac{1}{kT}\sum\limits_{\bf{pkq}}{{{f}_{0}}\left( \bf{p} \right){{f}_{0}}\left( \bf{k} \right)[1-{{f}_{0}}\left( {\bf{{p}'}} \right)][1-{{f}_{0}}\left( {\bf{{k}'}} \right)]
  W_{\bf{pk}, \bf{{p}'{k}'}}^{eh}{{\left[ \left( {{\Phi }_{p}}-{{\Phi }_{k}} \right)-\left( {{\Phi }_{{{p}'}}}-{{\Phi }_{{{k}'}}} \right) \right]}^{2}} }
\end{gather}

Above, we have introduced the Fermi golden rule scattering probabilities in unit time, given by:

\begin{gather}
\label{eq-w-ei}
W_{\bf{pp'}}^{ei}=\frac{2\pi}{\hbar }n_{\rm imp}{{\left| V\left( \bf{q} \right) \right|}^{2}}{{\left| M_{\bf{p{p}'}}^{++} \right|}^{2}}\delta \left( {{\varepsilon }_{\bf{p}}}-{{\varepsilon }_{\bf{p'}}} \right), \\
\label{eq-w-ee}
W_{\bf{pk}, \bf{{p}'{k}'}}^{ee}=\frac{2\pi }{\hbar }N{{\left| V\left( \bf{q} \right) \right|}^{2}}{{\left| M_{\bf{p{p}'}}^{++} \right|}^{2}}{{\left| M_{\bf{k{k}'}}^{++} \right|}^{2}}\delta \left( {{\varepsilon }_{\bf{p}}}+{{\varepsilon }_{\bf{k}}}-{{\varepsilon }_{{\bf{{p}'}}}}-{{\varepsilon }_{{\bf{{k}'}}}} \right), \\ 
\label{eq-w-eh}
 W_{\bf{pk}, \bf{{p}'{k}'}}^{eh}=\frac{2\pi }{\hbar }N\left[ {\left| V\left( \bf{q} \right) \right|}^{2}{{\left| M_{\bf{p{p}'}}^{--} \right|}^{2}}{{\left| M_{\bf{k{k}'}}^{++} \right|}^{2}}+{\left| V\left( \bf{p} +\bf{k} \right) \right|}^{2}{{\left| M_{\bf{pk}}^{+-} \right|}^{2}}{{\left| M_{\bf{p'k'}}^{+-} \right|}^{2}} \right]
 \delta \left( {{\varepsilon }_{\bf{p}}}+{{\varepsilon }_{\bf{k}}}-{{\varepsilon }_{{\bf{{p}'}}}}-{{\varepsilon }_{{\bf{{k}'}}}} \right) 
\end{gather}
where $N=4$ is the degeneracy factor, $\varepsilon_p = (E_g^2/4 + p^2v_0^2)^{1/2}$ is the energy spectrum in the gapped graphene, $V({\bf q}) = 2 \pi e^2/\kappa|{\bf q}|$ is the Fourier-transformed Coulomb potential, $\kappa$ is the background dielectric constant, $M^{ss'}_{\bf pp'}$ is the overlap of chiral wave functions between bands $s$ and $s'$ ($s=+1$ for the conduction and $s=-1$ for the valence band, respectively)
\begin{equation}
    |M^{ss'}_{\bf pp'}|^2 = \frac{\varepsilon_p   \varepsilon_{p'} + (E_g/2)^2+ ss' (v_0{\bf p} \cdot v_0 {\bf  p'})}{2 \varepsilon_p   \varepsilon_{p'} } .
\end{equation}

Two fundamental differences between effects of e-e and e-h collision integrals should be mentioned at that stage. First, there's a difference in signs with which the function $\Phi_{\bf p}$ enters the expressions (\ref{eq-q-ee}) and (\ref{eq-q-eh}). It stems from the fact that the entropy production rate ${\mathcal Q}$ is finite only when collisions change the electric current; and the quadratic-in-$\Phi$ expression in square brackets of (\ref{eq-q-ee}) and (\ref{eq-q-eh}) can be associated with the collision-induced change in electric current. Naturally, the current carried by two particles with momenta ${\bf p}$ and ${\bf k}$ depends on their charge, which explains the difference of collision integrals for e-e and e-h scattering.

Another difference between e-e and e-h collisions lies in the presence of an 'annihilation scattering', where electron and hole collide with virtual photon production, and subsequently yield yet another pair. Such scattering is represented by the second term in square brackets of (\ref{eq-w-eh}). In all other aspect, electron and hole collisions become identical if we neglect the exchange effects. The latter appear to be numerically small for quite a large number of particle 'sorts' $N=4$.

Our trial distribution function is selected as
\begin{equation}
\label{eq-trial}
\Phi_{\bf p} = \tau e ({\bf v_p} \cdot { \bf E})
\end{equation}
where $\tau$ is the parameter subjected to the optimization having the meaning of transport relaxation time, and ${\bf v_p} = \partial \varepsilon_{\bf p}/\partial {\bf p}$ is the electron velocity. Such choice of $\Phi_{\bf p}$ is important to reproduce the finite resistivity of gapless graphene. Other forms lead to the log-divergent collision integral due to the prolonged interaction of carriers with collinear momenta~\cite{Fritz2008}. In the gapped case, $\Phi_{\bf p}$ given by (\ref{eq-trial}) is not the only possible choice, but we stick to it for traceability of our result to the preceding studies.

The optimization of the entropy functional with respect to the scattering time yields the following result:
\begin{equation}
\label{eq-tau-sigma}
\tau^* = \frac{D}{C},\qquad \sigma_{eh} = D \tau^*,
\end{equation}
where $D$ is the Drude weight and $C = C_{ei} + C_{ee} + C_{eh}$ is net collision rate. Expressions for $D$ and $C$ are readily obtained from the entropy functional:
\begin{gather}
\label{eq-c-ei}
    C_{ei} = \frac{Ne^2}{2}\sum\limits_{\bf pp'}{\frac{\partial f_0}{\partial \varepsilon} ({\bf v_p} - {\bf v_{p'}})^2 W^{ei}_{\bf pp'}},\\
\label{eq-c-ee}
    C_{ee}=\frac{Ne^2}{8kT}\sum\limits_{\bf{pkq}}{
    {{f}_{0}}\left( \bf{p} \right){{f}_{0}}\left( \bf{k} \right)[1-{{f}_{0}}\left( {\bf{{p}'}} \right)][1-{{f}_{0}}\left( {\bf{{k}'}} \right)]
     W_{\bf{pk}, \bf{{p}'{k}'}}^{ee}{{\left[ \left( {{\bf{v}}_{\bf{p}}}+{{\bf{v}}_{\bf{k}}} \right)-\left( {{\bf{v}}_{{\bf{{p}'}}}}+{{\bf{v}}_{{\bf{{k}'}}}} \right) \right]}^{2}}},\\
\label{eq-c-eh}
    C_{eh}=\frac{Ne^2}{8kT}\sum\limits_{\bf{pkq}}{{{f}_{0}}\left( \bf{p} \right){{f}_{0}}\left( \bf{k} \right)[1-{{f}_{0}}\left( {\bf{{p}'}} \right)][1-{{f}_{0}}\left( {\bf{{k}'}} \right)]
    W_{\bf{pk}, \bf{{p}'{k}'}}^{eh}{{\left[ \left( {{\bf{v}}_{\bf{p}}}-{{\bf{v}}_{\bf{k}}} \right)-\left( {{\bf{v}}_{{\bf{{p}'}}}}-{{\bf{v}}_{{\bf{{k}'}}}} \right) \right]}^{2}} },\\
    D=-\frac{N e^2}{2}\sum\limits_{\mathbf{p}}{v_{p}^{2}}\frac{\partial {{f}_{0}}}{\partial \varepsilon }.
\end{gather}
\end{widetext}

Expressions for the electron-electron and electron-hole collision rates look very similar, and seem formally of the same order of magnitude. However, in parabolic gap case (realized at $E_g \gg kT$), the momentum conservation upon e-e collisions implies the conservation of total current by the virtue of proportionality ${\bf p} = m^* {\bf v_p}$. This feature is captured by Eq.~(\ref{eq-c-ee}), where the velocity factor in the square brackets is exactly zero if ${\bf p} = m^* {\bf v_p}$. Even in the gapless case, where proportionality between velocity and momentum does not hold, e-e collisions make a {\it numerically small} contribution to resistivity, compared to the e-h processes~\cite{Svintsov_eh_pumped}.

Expressions (\ref{eq-trial} - \ref{eq-c-eh}) are the central results of our paper and, in principle, enable the direct numerical evaluation of conductivity for the arbitrary value of the gap. For numerical purposes, it is convenient to eliminate the energy delta-functions by introducing the extra transferred energy variable $\omega$
\begin{multline}
\label{eq-delta-f}
 \delta \left( {{\varepsilon }_{\bf{p}}}+{{\varepsilon }_{\bf{k}}}-{{\varepsilon }_{{\bf{{p}'}}}}-{\varepsilon_{{\bf{k'}}}} \right) =\\
 \int{d\omega  \delta \left( \omega - {{\varepsilon }_{\bf{p}}}+{\varepsilon_{\bf p - \bf q}}\right) \delta \left( \omega - {{\varepsilon }_{\bf k + \bf q}}+\varepsilon_{\bf{k}} \right)},
\end{multline}
and integrate over the angles $\theta_{\bf pq}$ and $\theta_{\bf kq}$ analytically. Further simplifications are possible only in the limit of large gaps $E_g \gg kT$ and are described in Appendix A.

\section{Gap-dependent conductivity limited by carrier-carrier scattering}
We start our inspection of collision frequencies $\tau^{-1}$ and resistivity $\rho = \sigma^{-1}$ from the case of pristine graphene, i.e. neglecting the electron-impurity collisions. Such problem has only two dimensionless parameters: the coupling constant $\alpha_c = e^2/\kappa\hbar v_0$ and the normalized gap $E_g/kT$. Within the Born approximation to carrier-carrier scattering, the collision frequency and resistivity appear proportional to the coupling constant squared. Restoring the dimensionality of the collision rate and resistivity, we are always able to present them in the form:
\begin{gather}
    \tau_{eh}^{-1} = \alpha_c^2 \frac{kT}{\hbar} \tilde{\nu }\left(\frac{E_g}{kT}\right) ,\\
    \sigma_{eh} = \frac{e^2}{\hbar} \alpha_c^{-2}  \tilde{\sigma}\left(\frac{E_g}{kT}\right),
\end{gather}
where $\tilde{\nu }$ and $\tilde{\sigma}$ are the dimensionless collision frequency and the dimensionless conductivity depending only on the normalized gap $E_g/kT$.

\begin{figure}[ht]
\center{\includegraphics[width=1\linewidth]{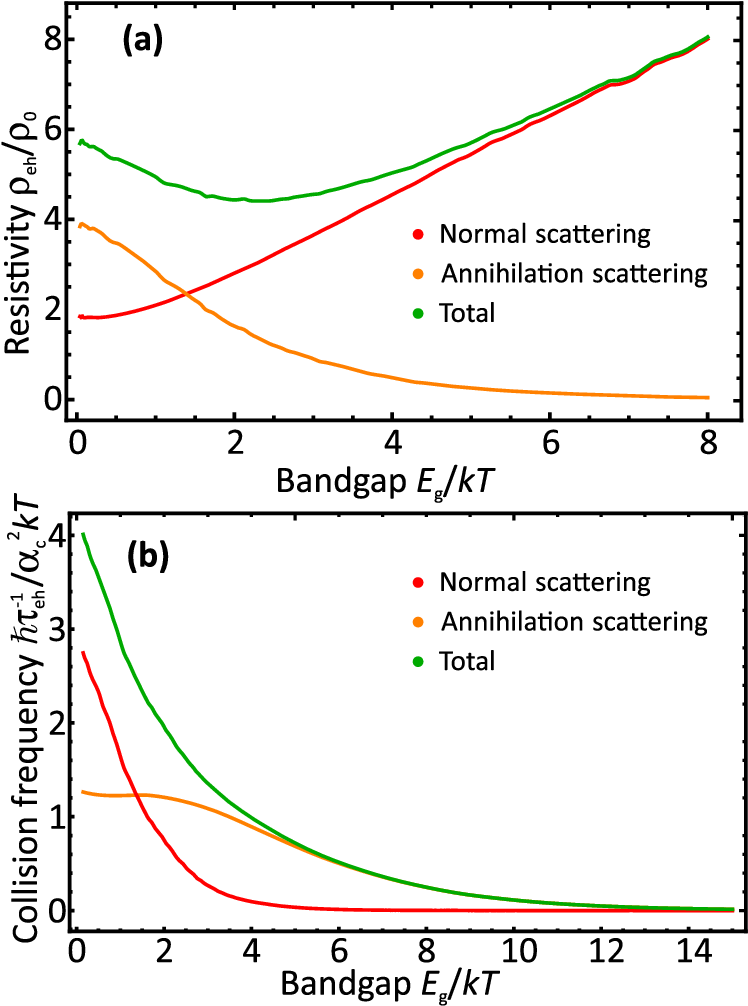}}
\caption{Electron-hole limited resistivity $\rho_{eh}$ (a) and associated collision frequency $\tau_{eh}$ (b) of pristine gapped graphene at charge neutrality point. The collision frequency is scaled by 'thermal frequency' $\alpha_c^2 kT/\hbar$, while the resisitivity is scaled by $\rho_0 = \alpha_c^2 h/2e^2$. The partial contributions of normal and annihilation-type electron-hole scattering are shown with red and orange curves, respectively}
\label{fig-pristine}
\end{figure}

The resulting dependences of collision rate and resistivity on $E_g/kT$ are shown in Fig.~\ref{fig-pristine}. The collision frequency $\tilde{\nu}$ decays monotonically with increasing the gap, which reflects the lack of thermally excited carriers to collide with in a non-degenerate semiconductor. It is possible to show that $\tilde{\nu} \propto e^{-E_g/2kT}$ at $E_g/kT \gg 1$. Still, the conductivity $\tilde\sigma_{eh}$ is scaling with gap non-exponentially. The reason is that Drude weight $D$ has a compensating exponential $D \propto e^{-E_g/2kT}$, which again reflects the exponentially small number of carriers in the band. The resulting dependence of $\tilde\sigma_{eh}$ on $E_g/kT$ appears to be hyperbolic at large values of gap, while the resistivity scales linearly, $\rho_{eh} \propto E_g/kT$. The latter scaling can be ascribed to the enhancement of effective mass in the Drude formula $\sigma_{eh} = ne^2\tau_{eh}/m^*$, provided the combination $n\tau$ is gap-independent.

Analyzing the intermediate-gap region, $E_g \sim kT$, it is instructive to split the electron-hole collision frequency into the contributions of 'normal' and 'annihilation-type' scattering.  The dependences of partial collision frequencies on $E_g/kT$ are both decaying, yet the annihilation collision frequency decays faster than the normal collision frequency. This fact is explained by approximate orthogonality of conduction and valence band states located in $kT$-window near the edges of conduction and valence bands. It is the states orthogonality which reduces the probability of annihilation, $|M^{+-}_{\bf p k}|^2 \ll 1$, and makes the annihilation-type collisions irrelevant to the conductivity of the large-gap semiconductor. Yet, the annihilation scattering is by no means small in the zero-gap state, and even appears stronger than the conventional scattering. 

Large frequency of annihilation-type collisions at $E_g=0$ and their rapid decay at $E_g \gg kT$ result in a non-trivial dependence of resistivity on band gap $E_g/kT$. At small induced gaps, the resistivity decays due to the rapid cancellation of annihilation-type scattering, and reaches a minimum at $E_g \approx 2.5 kT$. At larger values of $E_g$, the resistivity grows linearly due to the enhancement of effective mass. The presence of such {\it annihilation minimum} on the gap dependence of resistance is a natural hallmark of the carrier-carrier collision-limited transport.

To conclude the study of resitivity in pristine gapped graphene, we point to the main steps in deriving an approximate expression for resistivity at large band gaps. 
In that case, exchange electron-hole and electron-electron contributions to $C$ can be neglected. The Fermi distribution functions are reduced to the Boltzmann exponents, which simplifies the energy integration. The natural cutoff for the transferred momentum $q$ appears order of $E_g/v_0$ -- otherwise, the scattered electron cannot reside on the dispersion curve. The latter fact disables the appearance of the Coulomb logarithm~\cite{Landau_Collisions} in the expression for resistivity. Performing these steps, described in detail in Appendix A, we arrive at the conductivity of graphene with large induced gap in the form:
\begin{equation}
\label{eq-sigma-gapped}
   \sigma_{eh} (E_g \gg kT) = \frac{8}{\pi}\frac{e^2}{h} \alpha_c^{-2} \frac{kT}{E_g}.
\end{equation}
This value is very different numerically from the e-h limited conductivity of gapless graphene, obtained in Ref.~\onlinecite{Fritz2008} and reproduced in our calculations:
\begin{equation}
\label{eq-sigma-gapless}
   \sigma_{eh} (E_g = 0) =  0.76 \frac{e^2}{h} \alpha_c^{-2}.
\end{equation}
Despite numerical and functional differences between small- and large-gap asymptotics of $\sigma_{eh}$, both (\ref{eq-sigma-gapped}) and (\ref{eq-sigma-gapless}) can be presented in a similar form. This is achieved by introducing the 'running' coupling constant depending on the average thermal carrier velocity
\begin{equation}
\label{eq-alpha-running}
    \overline{\alpha}_c^2 = \frac{e^4}{\kappa^2 \hbar^2 \langle v_p^2 \rangle}.
\end{equation}
Here $\langle v_p^2 \rangle = v_0^2$ in the gapless state ($\overline{\alpha}_c \rightarrow \alpha_c$), and $\langle v_p^2 \rangle = 2 kT / m^* $ in the limit of large gap. With the notation (\ref{eq-alpha-running}), we present the large-gap conductivity as
\begin{equation}
   \sigma_{eh} (E_g \gg kT) = \frac{2}{\pi}\frac{e^2}{h} \overline{\alpha}_c^{-2} \approx 0.6 \frac{e^2}{h} \overline{\alpha}_c^{-2}.
\end{equation}

We may now speculate that the gap-dependent electron-hole collision-limited conductivity of variable-gap semiconductor is more universal than it was assumed previously~\cite{Tan_SciAdv_TunableSemimetal}. First, the scaling of $\sigma_{eh}$ with gap is non-exponential and much slower. Second, the conductivity can be expressed only via conductance quantum $e^2/h$ and the running coupling constant $\overline{\alpha}_c$, with a numerical prefactor very weakly depending on the gap.

\section{Observability of the electron-hole conductivity in disordered samples}

It is now tempting to compare the magnitudes of the carrier-carrier and carrier-impurity contributions to the resistivity in realistic disordered samples. Within the adopted variational approach, the contributions of these scattering channels to the collision frequency and resistivity are simply additive. From scaling considerations, the e-i collision frequency should be of the form:
\begin{equation}
    \tau^{-1}_{ei} = \alpha_c^2 \frac{n_{\rm imp}}{(kT/\hbar v_0)^2} \frac{kT}{\hbar} {\tilde \nu}_{ei} \left(\frac{E_g}{kT}\right).
\end{equation}
The normalized electron-impurity collision frequency depends weakly (non-exponentially) on the scaled band gap $E_g/kT$.

It becomes clear now that the electron-hole scattering-limited resistivity is dominant over impurity-limited if the hole density exceeds the impurity density. Of course, this conclusion is valid only at the charge neutrality point. With increasing the band gap, the number of thermally-excited carriers is reduced, while the density of impurities remains approximately constant. This makes us conclude that e-h collisions always become irrelevant at $E_g/kT \gg 1$.

\begin{figure}[ht]
\center{\includegraphics[width=1\linewidth]{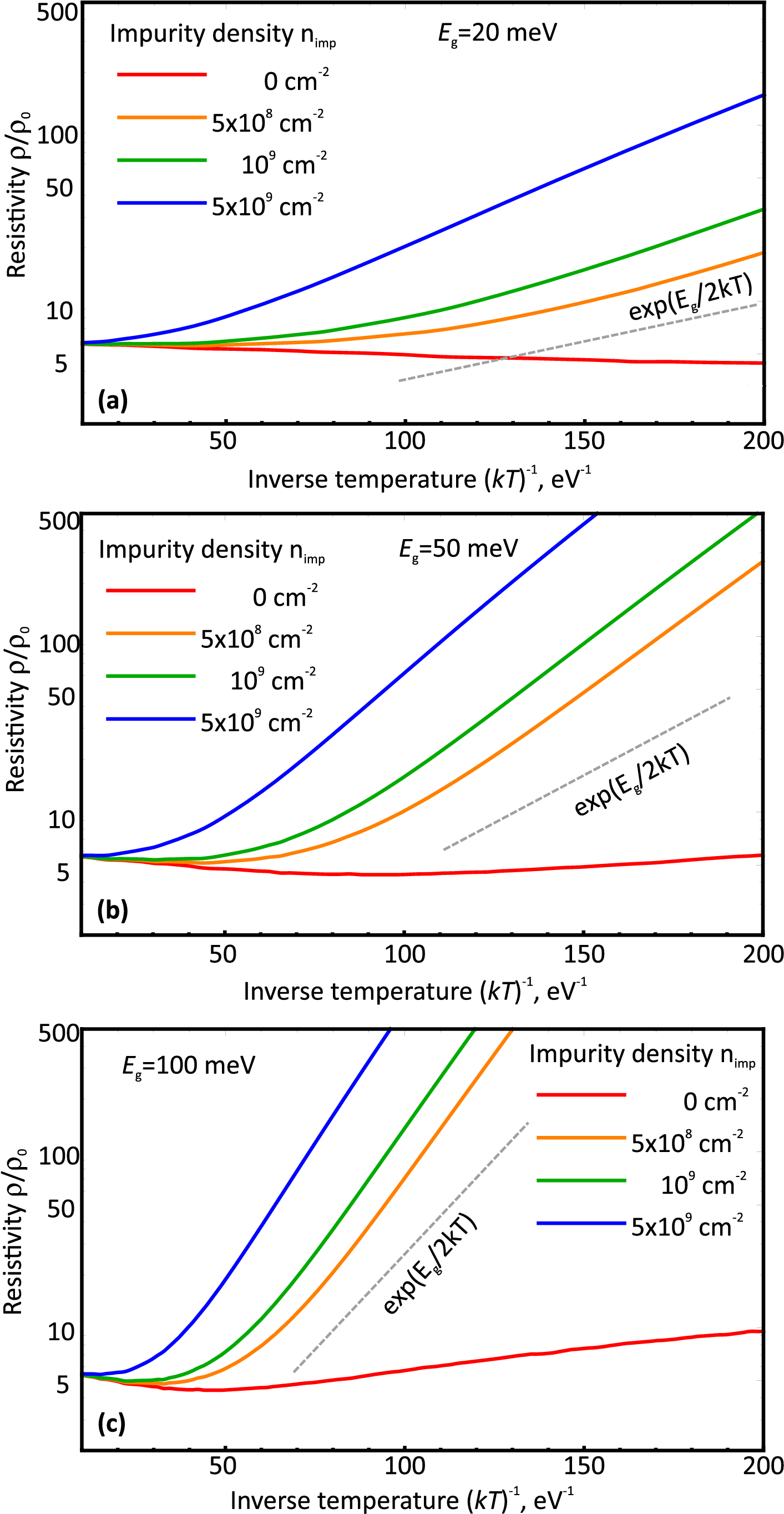}}
\caption{Temperature-dependent resistivity $\rho$ of gapped graphene under the combined action of electron-hole and impurity scattering for various values of band gap: $E_g = 20$ meV (a), $E_g = 50$ meV (b), and $E_g = 100$ meV (c). The resistivity is scaled by $\rho_0 =\alpha_c^2 h/2e^2$. Curves of different color correspond to different impurity densities $n_{\rm imp}$, red curve in each panel corresponds to the pristin graphene, gray dashed line is a guide for eye showing the Arrhenius behaviour}
\label{fig-impurities}
\end{figure}

The latter conclusion is illustrated in Fig.~\ref{fig-impurities}, where the total resistivity is shown as a function of inverse temperature at various impurity densities and band gaps. At the smallest band gap $E_g = 20$ meV (Fig.~\ref{fig-impurities}a), the total resistivity is dominated by e-h collisions in a wide temperature range starting from the highest temperatures to $(kT)^{-1} \approx 60$ eV$^{-1}$ ($T\approx 200$ K). With increasing the band gap to $E_g=50$ meV and then to $E_g=100$ meV (Fig. \ref{fig-impurities} b,c), we observe quite a rapid increase in total resistivity. The total resistivity deviates from electron-hole limited one already at nearly-room temperature, and becomes dominated by impurities. The Arrhenius law for impurity scattering is expectantly fulfilled: the logarithm of resistivity vs $T^{-1}$ becomes linear (see the gray dashed curve in Fig. \ref{fig-impurities}). We suggest that the seeming agreement between the exponential e-h conductivity scaling function $f(E_g/kT)$ of Ref.~\cite{Tan_SciAdv_TunableSemimetal} and experiment was due to the presence of residual impurities, which guaranteed the restoration of Arrhenius law. 

The density of impurities in gapped graphene should be pretty low to observe the effects of electron-hole scattering below the room temperature. For $E_g = 20$ meV and $n_{\rm imp} = 5\times10^{8}$ cm$^{-2}$, the e-h and e-i contributions to resistivity become equal to each other at $T\approx 90$ K. Such density of impurities is achievable in high-quality van der Waals heterostructures.

\section{Discussion and conclusions}

We have performed a systematic variational evaluation of resistivity due to electron-hole and electron-impurity scattering in gapped graphene. We have found that for pristine graphene, the resistivity at charge neutrality point scales as $\rho_{eh} \propto E_g/kT$. The scaling is non-exponential and violates the empirical Arrhenius law. This fact is related to simultaneous reduction in carrier density and collision frequency with increasing the band gap. A weak residual dependence of resistivity on the gap is due to the gap-dependent effective mass within the Dirac model. We expect that $\rho_{eh}$ would not depend on $E_g$ at all for semiconductors described by other Hamiltonians, where $E_g$ and $m^*$ are decoupled.

In this aspect, it's worth noting that gapped Dirac Hamiltonian does not provide an adequate description of electronic bands in graphene bilayer. Its spectrum remains parabolic at $E_g = 0$, and possesses a van Hove singularity at both band edges for finite gap~\cite{BandGap_GBL}. Evaluation of $\rho_{eh}$ for this kind of spectrum should be performed independently. Still, the cancellation of Arrhenius exponents in the expression for resistivity $\rho_{eh} = m^* \nu_{eh}/ne^2$ is a generic phenomenon independent of these spectral details.

Another important feature of gap-dependent resistivity $\rho_{eh}(E_g/kT)$ is the presence of minimum achieved at $E_g \approx 2.5 kT$. This minimum is provided by the competition of two opposite effects occurring upon gap induction: the enhancement of effective mass and reduction in probability of electron-hole annihilation scattering. Such an {\it annihilation minimum} is quite sensitive to the numerical values of the normal and annihilation scattering probabilities. Rigorously speaking, the minimum may not appear for other model potentials of electron-hole scattering, e.g. statically-screened or dynamically-screened Coulomb interaction. Still, the effects of screening should be small in the gapped case $E_g/kT \gg 1$, where the number of free carriers is low.

Our study was primarily devoted to the 2d systems with massive Dirac-type spectrum, where the maximum resistivity is achieved for equal and non-degenerate densities of electrons and holes. A different situation can emerge in semimetals with overlapping conduction and valence bands, where CNP is achieved for a degenerate electron-hole system. Such situation is realized in HgTe quantum wells above the critical thcikness, where e-h scattering-limited resistivity scales as $T^2$ \cite{Kvon_eh,Entin_eh}. Scattering between degenerate electrons and degenerate holes can also be realized in drag experiments~\cite{Bandurin_PRL_EHFriction} and in 2d semiconductors with interband population inversion~\cite{Svintsov_eh_pumped}.

All the above discussion concerned the resistivity of gapped graphene exactly at the charge neutrality point. Away from the CNP, the electron-hole collisions make a minor contribution to the resistivity due to the effect of Coulomb drag~\cite{Vyurkov_JETP,Zarenia_2dm}. More precisely, the majority carriers tend to drag the minority ones in the same direction, in spite of the opposite charges. The co-directional motion of electrons and holes reduces the strength of their mutual scattering. One may speculate that e-h scattering tends to nullify its own magnitude even for small density imbalance. An estimate of this imbalance can be obtained from the two-fluid hydrodynamic model~\cite{Svintsov_Hydro} and reads as $\delta n/ n_T \gtrsim \tau_{eh}/\tau_{ei}$. Definitely, a similar estimate could be obtained from a variational method using two different variational parameters $\tau_e$ and $\tau_h$. We refrain from these lengthy calculations due to the expected minor effect of e-h collisions away from the CNP.

Potential extensions of this work lie in the field of thermoelectric coefficients of gapped graphene in the electron-hole collision regime. The interest to thermoelectricity in such a system stems from (1) enhancement of Seebeck coefficient with gap induction even for impurity- and phonon-limited scattering (2) further enhancement of thermopower and reduction in thermal conduction in the hydrodynamic regime~\cite{Principi_Thermoelectrics_Hydro,Vignale_Thermal_conduction_hydro}. The combination of this factors can make clean gapped graphene a very strong candidate for thermoelectric photodetection~\cite{Titova_low_noise}.  

The work was supported by the grant \# 22-29-01034 of the Russian Science Foundation.

\appendix
\begin{widetext}
\section{Resistivity at large band gaps}
Throughout this section, we work in the units $\hbar = k = v_0 = 1$. The resistivity of gapped graphene at $E_g \gg T$ is limited by normal electron-hole scattering. Annihilation-type process can be neglected due to its small matrix element, while electron-electron scattering can be neglected due to the approximate conservation of current if carriers reside at the parabolic part of the spectrum. The contribution of normal e-h scattering to average collision rate $C$, which we denote by $C^{++}_{eh}$, can be presented in the form~\cite{Zarenia_2dm,Zheng_drag}
\begin{equation}
\label{eq-c-eh-normal}
    C_{eh}^{++}=
    \frac{e^2\pi^2}{T}
    \int q dq d\omega {
        |V(q)|^2 n_{\omega}[1+n_{\omega}]
    \left[
    {\Im}\Pi_0{\Im}\Pi_2 + {\Im}{\bf \Pi}_1{\Im}{\bf \Pi}_1
    \right],
    }
\end{equation}
where $n_\omega = [e^{\omega/T}-1]^{-1}$ are the Bose functions, and the 'generalized polarizabilities' $\Pi_n(q,\omega)$ are defined by
\begin{equation}
\label{eq-polarization-gen}
\Pi_{n}({\bf q}, \omega) = N \sum_{\bf p} 
{
\frac{f_0 (\varepsilon_p) - f_0(\varepsilon_{p-q})} {\varepsilon_p - \varepsilon_{p-q} - \hbar \omega + i \delta} |M^{++}_{{\bf p},{\bf p} - {\bf q}}|^2 ({\bf v_{p}} - {\bf v_{p-q}})^n.
}
\end{equation}
Representation (\ref{eq-c-eh-normal}) is achieved by introducing an extra energy variable $\omega$ with the aid of (\ref{eq-delta-f}) and developing the square of four particle velocities:
\begin{equation}
     [ {\bf v}_{\bf p }-{\bf v}_{\bf k }-{\bf{v}}_{{\bf{p - q}}}+{\bf v}_{\bf{k + q}}]^2 = ({\bf v_p} - {\bf v_{p-q}})^2    + ({\bf v_k} - {\bf v_{k+q}})^2 - 2({\bf v_p} - {\bf v_{p-q}})({\bf v_k} - {\bf v_{k+q}}).
\end{equation}

Evaluation of imaginary part of polarizability is achieved with Sokhotski rule
\begin{equation}
\label{eq-polarization}
\Im\Pi_0({\bf q}, \omega) = -
\pi N \sum_{\bf p}
{
\left[
f_0 (\varepsilon_{\bf p}) - f_0(\varepsilon_{\bf p-q})
\right]
\delta[\varepsilon_{\bf p} - \varepsilon_{\bf p-q} - \omega]
|M^{++}_{{\bf p},{\bf p} - {\bf q}}|^2.
}
\end{equation}
The angular integration is performed with the aid of delta function, which results in 
\begin{equation}
\label{eq-polarization-interated-angle}
\Im\Pi_0({\bf q}, \omega) 
= 
-\frac{N}{2 \pi}
\frac{1}{\sqrt{q^2-\omega^2}}
\int\limits_{E_{\min}}^\infty d\varepsilon
{
\left[
f_0 (\varepsilon+\frac{\omega}{2}) - f_0(\varepsilon - \frac{\omega}{2})
\right]
\frac{\varepsilon^2-\frac{q^2}{4}}{\sqrt{\varepsilon^2-E_{\min}^2}}.
}
\end{equation}
The minimum carrier energy at which the quantum $({\bf q},\omega)$ can be absorbed is denoted by $E_{\min}$:
\begin{equation}
    E_{\min}=\frac{q}{2}\sqrt{1+\frac{E_g^2}{q^2-\omega^2}}.
\end{equation}

Further integration over energies can be performed in the Boltzmann limit $E_g \gg T$. All slowly varying functions of energy $\varepsilon$ can be evaluated at $\varepsilon = E_{\min}$, while the remainder is evaluated exactly:
\begin{equation}
\label{eq-pi0-boltzmann}
\Im\Pi_0({\bf q}, \omega) 
= 
-\frac{\pi N}{(2 \pi)^2}
\frac{q^2 \sinh(\frac{\beta\omega}{2}) }{\sqrt{q^2-\omega^2}}
\sqrt{\pi} 
(B^2-1)
\frac{e^{-\beta B q/2}}{\sqrt{\beta B q}},
\end{equation}
where we have introduced the inverse temperature $\beta = T^{-1}$ and the dimensionless parameters
\begin{equation}
    B(q,\omega) = \sqrt{1+\frac{E_g^2}{q^2-\omega^2}},
    \quad
    l(q,\omega) = \frac{\omega}{q}
\end{equation}
Similar steps can be done to evaluate $\Im \Pi_{1,2}$ in the Boltzmann limit, which results in:
\begin{gather}
\label{eq-pi1-boltzmann}
{\Im}{\bf \Pi}_1 = 
2\pi \frac{N}{(2\pi)^2}
\sinh(\frac{\beta\omega}{2})
\sqrt{q^2-\omega^2} B
\sqrt{\pi}
\frac{B^2-1}{B^2-l^2}
\frac{e^{-\beta B q/2}}{\sqrt{\beta B q}}
\frac{{\bf q}}{q},\\
\label{eq-pi2-boltzmann}
{\Im}\Pi_2 = 
4\pi \frac{N}{(2\pi)^2} \sinh(\frac{\beta\omega}{2})
\sqrt{q^2  - \omega^2}
B^2
\sqrt{\pi}
\frac{B^2-1}{B^2-l^2}
\frac{1-l^2}{B^2-l^2}
\frac{e^{-\beta B q/2}}{\sqrt{\beta B q}}.
\end{gather}

Considerable cancellations appear after collecting the expressions (\ref{eq-pi0-boltzmann} - \ref{eq-pi2-boltzmann}) into $C^{++}_{eh}$:
\begin{equation}
\label{eq-ceh-boltzmann}
C_{eh}^{++}=
    \frac{\pi^3N^2 e^2\alpha_c^2}{8}
    \int dq d\omega
\exp\left[-\beta E_g q B(q,\omega) \right]
\frac{q^2 B(q,\omega) (q^2-\omega^2)}{
\left[
q^2+\frac{(q^2-\omega^2)^2}{E_g^2}
\right]^2}
\end{equation}
where $q$ and $\omega$ were normalized to the band gap $E_g$.

Let us inspect the function under the exponent $f(q) = q B (q, \omega)$. Considered as a function of $q$, it has a minimum value $f_{\min} = 1 + \omega$. Instead of $q$, we pass to the new variable 
\begin{equation}
\tau = q B (q, \omega).
\end{equation}
The dimensionless integral in the expression for $C^{++}_{eh}$ is now recast in the form
\begin{equation}
   I \equiv \int dq d\omega
\exp\left[-\beta E_g q B(q,\omega) \right]
\frac{q^2 B(q,\omega) (q^2-\omega^2)}{
\left[
q^2+\frac{(q^2-\omega^2)^2}{E_g^2}
\right]^2}
=  
2
\int\limits_{0}^{\infty} d{\omega}
\int\limits_{1 + \omega}^{\infty} d\tau
\frac{\tau^2}{(\tau^2-{\omega}^2)^2}
\frac{e^{-\beta E_g \tau}}{
    \sqrt{(\tau^2 - {\omega}^2 - 1)^2 - 4 {\omega}^2}
}.
\end{equation}
Integrating over $\tau$, we again proceed with the steepest descend method, i.e. we set $\tau = 1+\omega$ in all smooth functions under the integration sign. This results in  
\begin{multline}
I \approx
\frac{(\beta E_g)^2}{2}
\int\limits_{0}^{\infty} d{\omega}
\frac{(1+\omega)^2}{((1+\omega)^2-{\omega}^2)^2}
\frac{1}{2\sqrt{{\omega}+1}}
\int\limits_{1+\omega}^{\infty} d\tau
\frac{e^{-\beta E_g \tau}}{
    \sqrt{\tau^2 - (1+\omega)^2}
} = \\
= \frac{(\beta E_g)^2}{4} e^{-\beta E_g}
\int\limits_{0}^{\infty} d\omega
\frac{({\omega}+1)^{3/2}}{(1+2{\omega})^2}
K_0({\omega}\beta E_g)
\approx
\frac{\pi \beta E_g}{8} 
e^{-\beta E_g}.
\end{multline}
Hence, 
\begin{equation}
\label{eq-ceh-final}
    C_{eh}^{++}=
    \frac{N^2 e^2 \alpha^2 }{8}
    (\beta E_g)
    \frac{\pi}{8} e^{-\beta E_g}.
\end{equation}
The Drude weight $D$ is also simply evaluated with Boltzmann statistics
\begin{equation}
\label{eq-drude-boltzmann}
    D \approx \frac{N e^2}{2\pi \hbar^2} k T e^{- \beta E_g/2}.
\end{equation}
Substituting $C$ and $D$ in the Boltzmann limit [Eqs.~(\ref{eq-ceh-final}) and (\ref{eq-drude-boltzmann})] into the general expression for the conductivity (\ref{eq-tau-sigma}), we find the linear-in-$T$ scaling of conductivity (\ref{eq-sigma-gapped}).

\end{widetext}

\bibliography{Refs}

\end{document}